# Phone physics and the Gateway Arch: Fun with friends and physics at the AAPT Winter Meeting in St. Louis

David Marasco, Foothill College, Los Altos Hills, CA

Bree Barnett Dreyfuss, Amador Valley High School, Pleasanton, CA

As a famous landmark and feat of engineering, the Gateway Arch was a popular destination at the 2025 AAPT Winter Meeting in St. Louis. The visit to the observation deck of the Gateway Arch is unique, climbing the steps after exiting the small tram capsules and seeing a floor that continues to slope upward assures that you are in fact at the very top. Everyone in our group excitedly took pictures, pointing out local features like the Dred Scott Courthouse. There were many selfies at the pinnacle, and we discussed how to work them into future questions for our students. During our tram ride to the top observation deck of the arch, we lamented that we should have brought pendula to measure the acceleration due to gravity. You can take physics teachers out of the physics conference, but you apparently can't get us to stop talking about physics teaching. Recognizing that we had accelerometers on our phones we collected data on the descent. The authors wanted to collect more complete measurements and returned two days later to repeat the journey, the results of which we present here. For readers wishing to repeat with their students, or who want to apply more advanced data analysis techniques, the authors have made the raw data, our spreadsheets, and a teacher's guide available[1].

National Park Service tour guides explained the mechanics of the tram system as we awaited our tour. One of our guides gave a particularly detailed explanation like he knew there were a bunch of physics teachers in the audience who did actually want to know everything. The tram system (Figure 1) is described as a Ferris wheel that is like an escalator and an elevator. The journey begins with the tram capsules moving along the track at their top (like a Ferris wheel), then rotates to be connected at their side as they move up almost vertically (like an elevator) before continuing up towards the top of the arch at an angle (like an escalator)[2]. Thus the tram's motion was limited to the vertical and one horizontal direction.

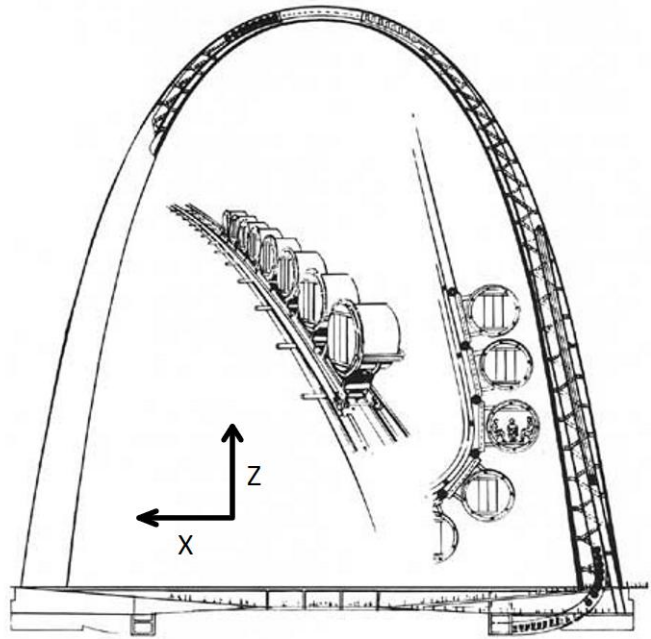

Figure 1. An architect's rendering of the tram system of the St. Louis Arch in different positions[3]. Coordinate axes added for reference

Previous authors used accelerometers to investigate the motion of elevators[4-7], therefore data were collected to be used with high school or first year college physics students to explore both relationships in kinematics and to apply them in real-life scenarios. As this was the intent, the analysis is presented at that level. While multiple phones were initially used, only one recorded both the ascent and descent data on that second trip to the Gateway Arch, which ultimately acted as an initial fitting constraint. Pressure sensor and accelerometer data were collected by the PhyPhox[8] app on a Samsung Galaxy S23. AAPT's Winter Meeting preceded the March 2025 publication of Yu's work on zeroing cell phone accelerometers[9]. Yu's methods would have greatly benefited our results, as much of our analysis concerned correcting for offsets in the accelerometer data.

The PhyPhox app has an "Elevator" experiment which presents altitude and velocity from the phone's barometer and also the Z-acceleration from the accelerometer. We followed this example, collecting both barometer and accelerometer data. The change in pressure is proportional to change in weight above the area:

$$\Delta P = (\rho\, g\, A\, \Delta h)/A \tag{1}$$

where P is the pressure, $\rho$ is the density of air, g is the acceleration due to gravity, A is the cross-sectional area of the air column and h is the height; the value in the numerator of the ratio is a change in weight. This approach using pressure (Method 1) gave a maximum displacement of 198 meters on the height graph in Figure 2. With the knowledge that the height of the Gateway Arch is 192 meters[3], counting pixels on the diagram in Figure 1 gives a tram car displacement of 203∓2 meters, depending on choice of tram. The PhyPhox Elevator code implements a barometric formula (Method 2) relating pressure to altitude[10,11], and generates results within 50 centimeters of Method 1. The five meter difference between the height found via Method 1 and by inspection of Figure 1 is reasonable. In addition to the uncertainty on the displacement from the diagram, the pressure measurement assumed an air density of 1.225 kg/m³, which is the value for dry air at 15°C [12]. That temperature is not out of the question as it was well below freezing outside, and the authors wore their jackets even inside. However, if the air was 20°C the density drops to 1.204 kg/m³, which would give a height difference of 3.5 meters.

With an established height profile, the next step was to take the Z-axis accelerometer readings and use Euler's Method, numerically integrating twice to find the displacement (Method 3). For each row in the spreadsheet[1] the velocity was calculated by summing the previous velocity with the accelerometer value multiplied by the incremental time step. This was repeated for the displacement, using the velocity. Note that this double-integration is not a built-in feature of PhyPhox, in fact their documentation points out that "the noise of the sensors brings the results to absolutely unreasonable values within a short time."[13] Kinser finds displacements using this method can be off by more than a factor of 2 for simple elevators with near-constant accelerations[4]. This was compounded by the fact that the Z-axis accelerometer was recording a value of roughly 9.8 $\frac{m}{s^2}$, corresponding to the acceleration due to gravity, so an offset needed to be subtracted. Since the round trip of the tram started and finished in the same vertical position, the offset to the acceleration was varied from the average sensor value of 9.8234 $\frac{m}{s^2}$ until the final position matched the initial position. The offset that satisfied those boundary conditions was 9.8263 $\frac{m}{s^2}$. This revealed the height profile in Figure 2 shown in blue, with a maximum displacement of 262 meters. Note that the error in displacement from accelerometer offset grows quadratically with time due to the numerical double integration. The roughly 60 meter overshoot over the 450 second long experiment would be a mere three centimeters in a ten second experiment, which is more encouraging for people wishing to use the accelerometers on their phones for shorter exercises.

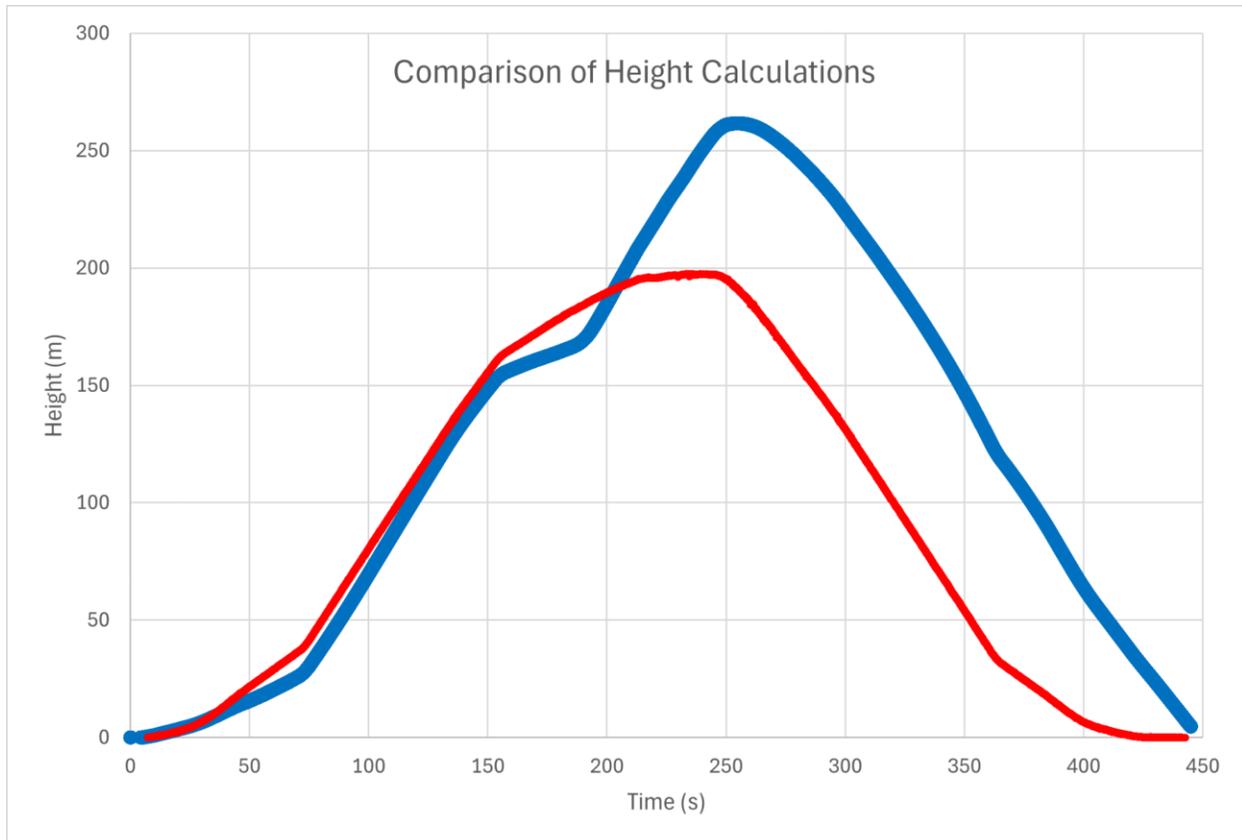

*Figure 2. Height vs time calculations: Method 1 (pressure) results in red, Method 3 (Euler) results in blue.*

While the peak displacement values between the two methods differ by 36%, the kink at roughly 150 seconds is consistent in Figure 2. The tram slows down its vertical velocity at that point in both Method 1 and Method 3 analyses. However, in the pressure treatment (Method 1) the velocity slows down again at 215 seconds then stops, whereas the accelerometer shows the tram speeding up to approximately the previous velocity at roughly 190 seconds. On the trip down, the barometer data suggest that the tram gently slows to a stop, whereas the accelerometer has it moving at constant vertical speed, moving at a negative velocity up until the end of travel. A visual comparison of velocity graphs (Figures 3 & 4) at this point is instructive. Note that both are generally consistent with the top speed of $1.7 \frac{m}{s}$ claimed by the tour guide.

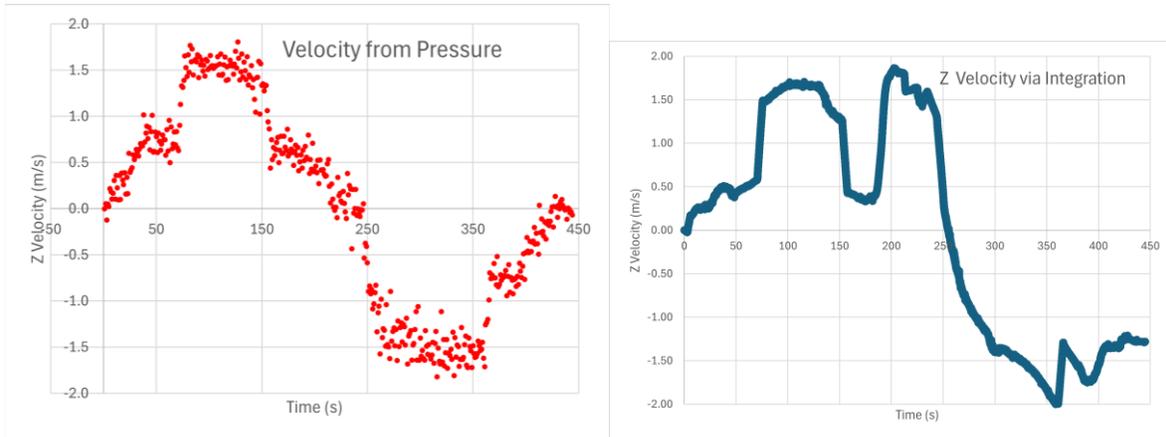

*Figures 3 & 4. Z-Velocity determined via Method 1 (pressure) and Method 3 (Euler)*

The velocity graph via integration (Figure 4) has a positive but decreasing velocity in the Z-direction when the tram starts moving downward. This highlights a common student misconception, that the tram is moving down "when the line is moving down." Of course, the tram is moving downward only when the Z-direction velocity is negative. As noted by Ustin and Cetin[14], points where the object in question is known to be at rest should be used to reset the velocity to zero, to rollback accumulated error. Hence the data set is split into two parts, with each segment assigned a different acceleration sensor offset, using the information that the phone is at rest when data collection is restarted on the way down. This approach (Method 4) generates the following two height plots (Figures 5 & 6). The offset moving upward is $9.826 \frac{m}{s^2}$ and downward $9.812 \frac{m}{s^2}$. The estimated maximum height by accelerometer data is slightly improved to 251 meters.

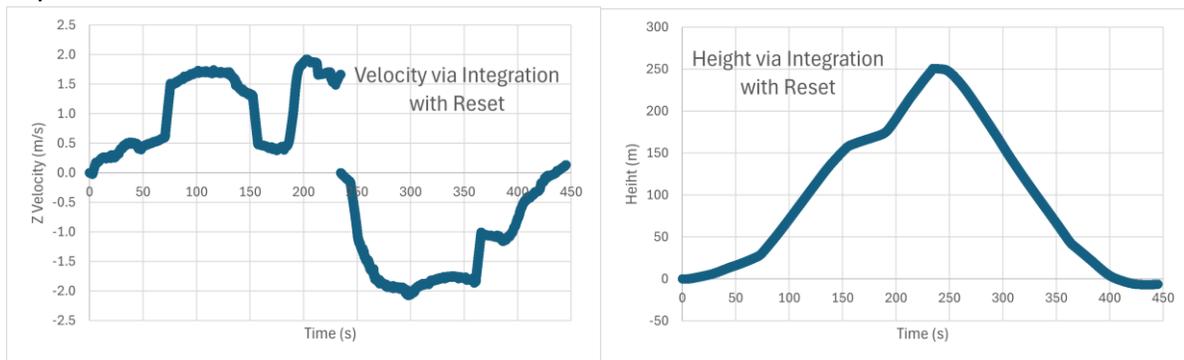

*Figures. 5 & 6. Z-direction Velocity & Height via Method 4 (Euler with reset)*

Consistent with Monteiro and Marti, the barometer model is a stronger representation than integrated acceleration[6], the displacement matches better with the known height of the Gateway Arch. This may be because the tram changes its orientation as it travels, a necessity of the shape of the structure. It does this in a similar fashion to a Ferris wheel, which introduces rocking, so while the Z-axis is generally pointing upward, the orientation of the phone's and the Earth's Z-axes is not consistent, which exacerbates the previously stated issues with two numerical integrations on noisy sensor data. Even a one-degree offset in the Z-axis would change the measured acceleration by $0.0015 \frac{m}{s^2}$ for that time interval, comparable to some of the offset adjustments seen previously. It is likely that some angular deviations were considerably larger.

The results for the horizontal directions are problematic. No combination of acceleration offsets for the X-direction generate results that are truly satisfying in both velocity and position for round trip data (Figures 7 & 8). While the signs of concavities of the graph are symmetric with travel, almost nothing else is. The asymmetry in travel times (235 seconds to go up, roughly 200 seconds to go down) may lead to different noise/vibration environments, and very different accelerator sensor offsets (0.0278m/s² going up, 0.105m/s² going down). That being said, the scale of the distances is within reason. The Gateway Arch is 192 meters wide at its base, so the leg-to-center distance is 96 meters, whereas the extremes in the X-displacement using Method 4 differ by roughly 90 meters. The tram starts from a position inside the curve of the structure, which accounts for the initial negative displacement. The Y axis should give no motion, but yields a wild maximum displacement result of over a kilometer. This may point to issues with integrating a variable that is dominated by noise.

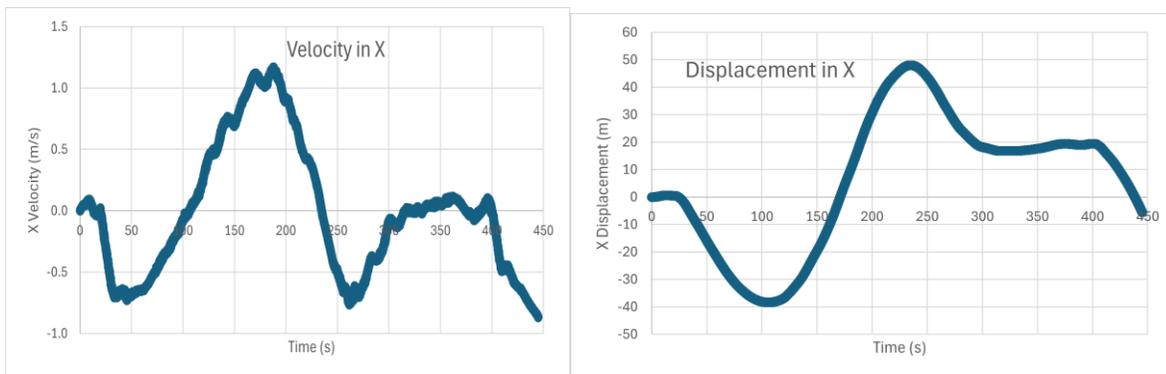

Figures. 7 & 8. Calculated horizontal velocity and displacement to time via Method 4 (Euler with reset).

The shape of the Gateway Arch is that of a weighted catenary[15]. Plotting vertical displacement against horizontal displacement on the way up parameterized by time does not result in a catenary, which is unsurprising given what was previously known about the displacement via integration data. That being said, the graph does catch an important detail, from the visitor's hall the tram first moves away from the center of the Gateway Arch before moving upwards and back towards the center. Graphing the calculated horizontal displacement with the height derived from pressure (Method 1) better matches the arch. The displacement profiles are compared to the path of the tram in the Gateway Arch in Figure 9.

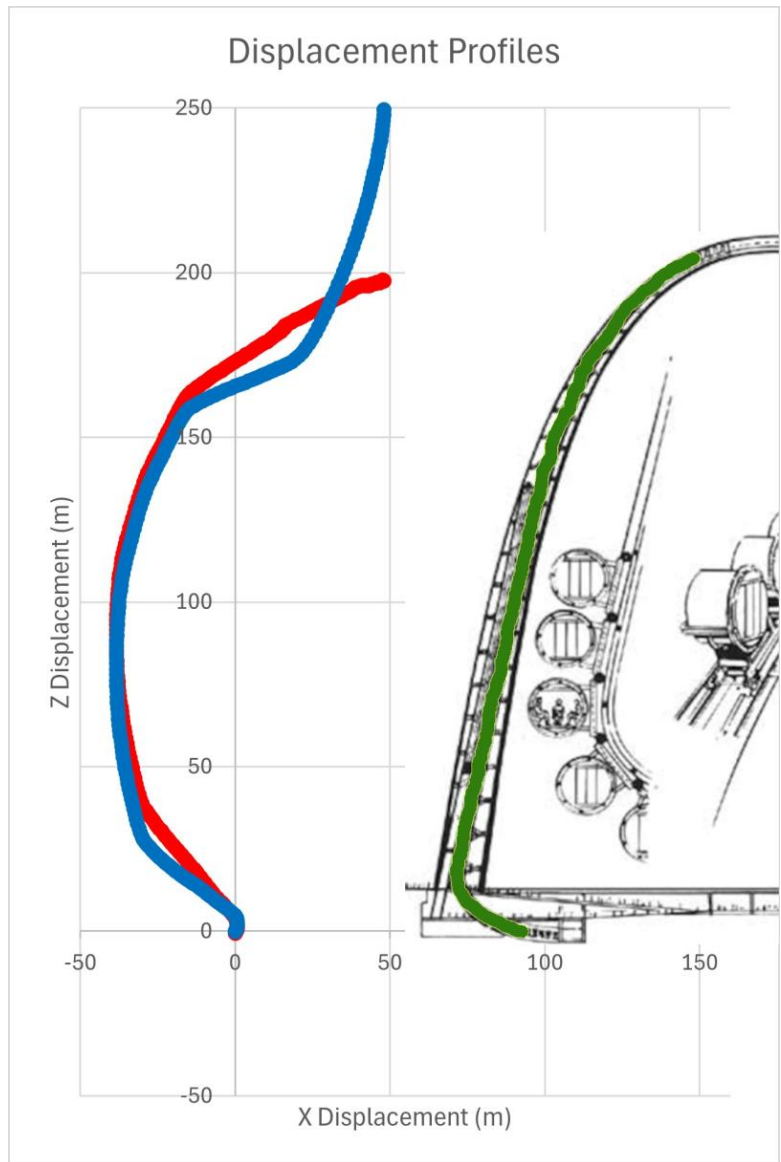

*Figure. 9. Blue curve is from integrating both directions from accelerometer data (Method 3), red is horizontal integration (Method 3) and height from pressure (Method 1), green is the path of the tram.*

While the results may not match textbook examples, that these data were collected on a whim to satisfy curiosity raises student interest levels. If presenting the raw data to students to analyze, asking them to first think about what they would expect to see encourages critical thinking when the results aren't perfect. Understanding the flaws and limitations of their tools is an important skill scientists use every day. And of course, the best lesson for our students may be that physics is everywhere, even when playing tourist (Figure. 10).

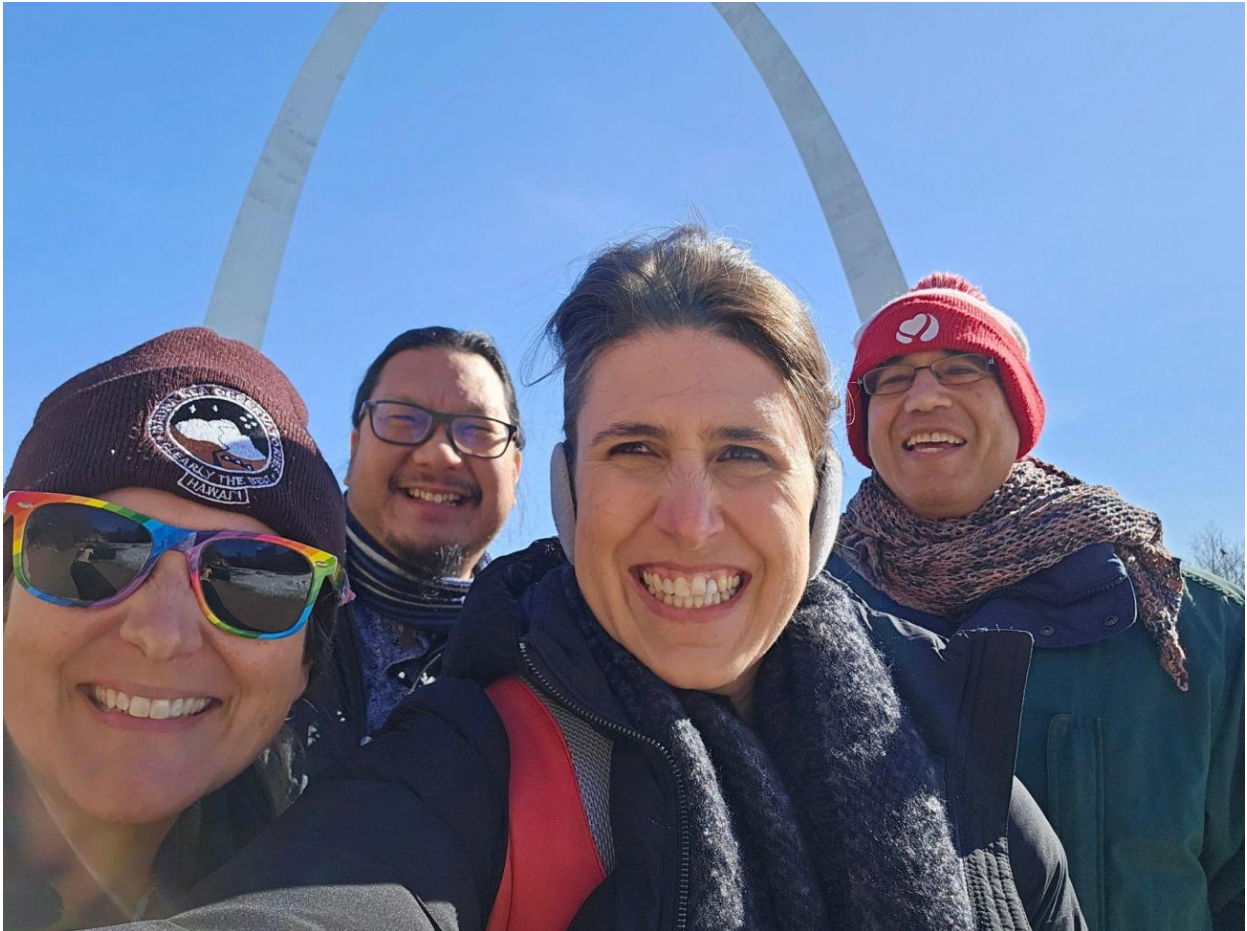
*Figure. 10. AAPT attendees Bryn Bishop and Jay Kurima with authors Bree Barnett Dreyfuss and David Marasco.*

Acknowledgments
The authors would like to thank Elissa Levy, Jay Kurima, and Bryn Bishop for companionship and encouragement, and Dan Burns for his suggestions on data collection for our second visit to the Gateway Arch.